\documentclass[prd,preprint,nofootinbib]{revtex4}

\usepackage{graphicx, cancel}
\usepackage{epsf}
\usepackage{amsmath}
\usepackage{graphics}
\usepackage{amsfonts}
\usepackage{amssymb}
\usepackage{latexsym}
\usepackage{color}
\input{colordvi.tex}

\begin{document}
\renewcommand{\thefootnote}{\#\arabic{footnote}}
\newcommand{\rem}[1]{{\bf [#1]}}

\newcommand{\gsim}{ \mathop{}_ 
{\textstyle \sim}^{\textstyle >} }
\newcommand{\lsim}{ \mathop{}_ 
{\textstyle \sim}^{\textstyle <} }
\newcommand{\vev}[1]{ \left\langle {#1}  
\right\rangle }

\newcommand{\bear}{\begin{array}}  \newcommand 
{\eear}{\end{array}}
\newcommand{\bea}{\begin{eqnarray}}   
\newcommand{\eea}{\end{eqnarray}}
\newcommand{\beq}{\begin{equation}}   
\newcommand{\eeq}{\end{equation}}
\newcommand{\bef}{\begin{figure}}  \newcommand 
{\eef}{\end{figure}}
\newcommand{\bec}{\begin{center}}  \newcommand 
{\eec}{\end{center}}
\newcommand{\non}{\nonumber}  \newcommand 
{\eqn}[1]{\beq {#1}\eeq}
\newcommand{\la}{\left\langle}  
\newcommand{\ra}{\right\rangle}
\newcommand{\ds}{\displaystyle}

\def\SEC#1{Sec.~\ref{#1}}
\def\FIG#1{Fig.~\ref{#1}}
\def\EQ#1{Eq.~(\ref{#1})}
\def\EQS#1{Eqs.~(\ref{#1})}
\def\lrf#1#2{ \left(\frac{#1}{#2}\right)}
\def\lrfp#1#2#3{ \left(\frac{#1}{#2} 
\right)^{#3}}
\def\GEV#1{10^{#1}{\rm\,GeV}}
\def\MEV#1{10^{#1}{\rm\,MeV}}
\def\KEV#1{10^{#1}{\rm\,keV}}

\def\lrf#1#2{ \left(\frac{#1}{#2}\right)}
\def\lrfp#1#2#3{ \left(\frac{#1}{#2} 
\right)^{#3}}

\def\A{\mathcal{A}}
\renewcommand{\thefootnote}{\alph{footnote}}

\renewcommand{\thefootnote}{\fnsymbol{footnote}}
\preprint{IPMU 08-0015}
\title{Non-Gaussianity from Symmetry}
\renewcommand{\thefootnote}{\alph{footnote}}

\author{Teruaki Suyama$^{1}$ and Fuminobu Takahashi$^{2}$}

\affiliation{
${}^1$Institute for Cosmic Ray Research, University of Tokyo,
Kashiwa 277 8582, Japan,\\
${}^2$ Institute for the Physics and Mathematics of the Universe,
   University of Tokyo, Chiba 277-8568, Japan
   }

\begin{abstract}
\noindent
We point out that a light scalar field fluctuating around a
symmetry-enhaced point can generate large non-Gaussianity in density
fluctuations. We name such a particle as an ``{\it ungaussiton}", a
scalar field dominantly produced by the quantum fluctuations,
generating sizable non-Gaussianity in the density fluctuations.  We
derive a consistency relation between the bispectrum and the
trispectrum, $\tau_{\rm NL} \sim 10^3 \cdot f_{\rm NL}^{4/3}$, which
can be extended to arbitrary high order correlation functions. If such
a relation is confirmed by future observations, it will strongly
support this mechanism.
\end{abstract}
\maketitle

%%%%%%%%%%%%
\section{Introduction}
\label{sec:1}
%%%%%%%%%%%%

The concept of symmetry has been a guiding principle in modern
physics.  The structure of the standard model (SM) is dictated by the
SM gauge symmetries, $SU(3)_C \times SU(2)_L \times U(1)_Y$.  In the
string theory, gauge symmetries as well as discrete global symmetries
are ubiquitous~\cite{string}.

One of the plausible candidates for the theory beyond SM is
supersymmetry (SUSY), which relates a particle to a superparticle with
different statistics.  In particular, a superpartner of a fermion
field with spin $1/2$ is a scalar field.  Therefore there are perhaps
many fundamental (or composite) scalar fields in nature, and most of
them may be charged under some symmetries.

A scalar field that is singlet under any symmetries does not possess a
special point in its field space, and so, the minimum of the potential
may vary in the evolution of the universe.  The change of the minimum
generically induces coherent oscillations of the scalar field. If the
scalar field has only interactions suppressed by the Planck scale,
$M_P (=2.4 \times 10^{18}{\rm\, GeV})$, it might induce notorious
cosmological moduli problem~\cite{ModuliProblem}, which has recently
turned out to be more acute than previously
thought~\cite{MGP,MGP2,Endo:2006tf}.  That is to say, the modulus
decay generically produces too many gravitinos, which either spoil the
success of the big bang nucleosynthesis or exceed the observed dark
matter abundance. The moduli problem and the moduli-induced gravitino
problem can be ameliorated if the initial displacement is somehow
suppressed. This, however, is difficult to achieve unless the modulus
field is charged under some symmetries.

On the other hand, a scalar field charged under some symmetries has a
special point in its field space~\footnote{One exception is an axion
  with a shift symmetry, because the axion does not have a
  symmetry-enhanced point. We will make a comment on this case
  later.}, where the symmetries are enhanced and the scalar potential
has its extremum. If the symmetry-enhanced point is the (local)
minimum of the potential, it is quite likely that the scalar field
sits at the point from the beginning of the universe and continues to
sit there~\cite{Dine:1998qr,
  Kofman:2004yc,Watson:2004aq,Cremonini:2006sx, Greene:2007sa}.  Such
a scalar field apparently does not affect the subsequent evolution of
the universe, since its energy density vanishes, at least
classically. At the quantum level, however, this may not hold as we
describe below.

The inflation is now strongly suggested by the recent observations of
the cosmic microwave background (CMB). During inflation, the inflaton
acquires quantum fluctuations, which become the seed for the present
structures of the universe.  Interestingly, any scalar fields of
masses smaller than the expansion rate during inflation, $H_I$,
similarly acquire quantum fluctuations of $O(H_I)$~\cite{GPP}.

We point out that such light scalar fields fluctuating around the
symmetry-enhanced point necessarily induce non-Gaussianity at some
level in the observed CMB power spectrum~\footnote{As will be
  discussed later, the symmetries are not necessarily exact. Our
  arguments remain intact as long as the deviation from the potential
  minimum is small (or comparable) relative to the Hubble parameter
  during inflation. }.  In other words, the scalar fields whose energy
classically vanishes must have left their traces in the CMB; those
scalar fields are produced by the purely quantum fluctuations, making
the fluctuations in their energy density deviate from the Gaussian
distribution.  We call such a scalar particle an ``{\it ungaussiton}",
because their energy density exhibits strong
non-Gaussianity~\cite{Linde:1996gt,Durrer:1998sv}.  Our definition of
the ungaussiton is {\it an elementary or composite scalar field
  dominantly produced by the quantum fluctuations, generating sizable
  non-Gaussianity in the density fluctuations. }  Note that the
definition itself is not related to the symmetry. The presence of the
symmetry is one of the ways to naturally make a scalar field to be an
ungaussiton.

 If there are indeed many scalar fields with symmetries in nature,
 some of them may be light during inflation.  Then they can be
 ungaussitons.  In this way we expect that the presence of
 non-Gaussianity might be quite generic in our vacuum.  Interestingly,
 it has been recently reported that large non-Gaussianity was detected
 by the analysis on the WMAP 3yr data~\cite{Yadav:2007yy}.  The latest
 WMAP 5yr data seem to have the same tendency, although the vanishing
 non-Gaussianity is allowed within 95\% C.L.~\cite{Komatsu:2008hk}.
 Those hints on the non-Gaussianity, or its future detection, may be
 originated from such particles, i.e., ungaussitons.

In this paper we study the ungaussiton mechanism and write down a
condition for the non-Gaussianity to become large enough to be
observed by the ongoing and planned observation such as the WMAP and
Planck satellites.  Moreover, we derive a consistency relation between
the bispectrum and the trispectrum, which, if confirmed, will be a
smoking gun of the ungaussiton mechanism. Importantly, the bispectrum
and the trispectrum predicted in the ungaussiton scenario exhibit a
specific dependence on the scale; both get enhanced as one goes to
smaller scales.  Such a feature on the scale dependence as well as the
consistency relation between the bispectrum and the trispectrum will
help us to distinguish our scenario from the others.
 
  Let us here comment on how the ungaussiton can remain around the
   minimum of the potential during inflation. As is well known, if
   inflation lasts sufficiently long, a light scalar field (an
   ungaussiton in our case) relaxes towards the Bunch-Davies (BD)
   distribution~\cite{Bunch:1978yq}. The BD distribution is such that
   the root mean square (RMS) of the field fluctuations on large
   scales is given by ${\cal O}(H_I^2/m)$, where $m$ is the effective
   mass of the ungaussiton during inflation.  If $m$ is much smaller
   than $H_I$, the BD distribution tells us that the ungaussiton takes
   a value of ${\cal O}( H_I^2/m) \gg H_I$ in most regions, while
   there is a small portion of the whole universe where the field
   value is of ${\cal O}(H_I)$. If our observable universe happens to
   be inside such small region, the spatial average of the ungaussiton
   over the observable universe can be very close to the origin. This
   may sound a fine-tuning, but it may not. There may be anthropic
   arguments to select such regions, although there is no
   justification for such reasoning at this moment.  There are other
   ways to keep the ungaussiton near the origin.  One possibility is
   that the mass $m$ is not much smaller than $H_I$.  Then the RMS
   values of the ungaussiton is not that large compared to $H_I$,
   which makes it easy to find such regions that the field value is of
   $O(H_I)$ without fine-tunings.  Another is that the last inflation
   was not long enough.  For instance, the $e$-foldings of $50 \sim
   60$ is too short for the field to reach the BD distribution.  The
   ungaussiton then stays near the minimum of the potential throughout
   the inflation, if the ungaussiton dynamically relaxed to the origin
   before the inflation. We will come back to this issue in
   Sec.~\ref{sec:4}.

%%%%%%%%%%%%%%%%%%%%%%%%%%%%%%%
\section{Non-Gaussianity from an ungaussiton}
\label{sec:2}
%%%%%%%%%%%%%%%%%%%%%%%%%%%%%%%

Let us consider a scalar field, $\sigma$, which is charged under some
symmetries~\footnote{ It is easy to extend our results to a case of
  the multiple scalars.  }. We set the origin of $\sigma$ to be the
symmetry-enhanced point, which is assumed to be the (local) minimum of
the potential. If it is a meta-stable vacuum, the life time of the
vacuum is assumed to be much longer than the present age of the
universe. The potential around the origin can be obtained by expanding
the scalar potential in terms of $\sigma$. We assume that the
potential is well approximated by the quadratic potential~\footnote{
One can also consider a case that a quartic potential dominates over
the quadratic one. The resultant non-Gaussianity tends to be smaller
in this case.
},
and that it is valid at least up to $\sigma = O(H_I)$:
\beq
U(\sigma) \;=\; \frac{1}{2} m_{\sigma}^2 \sigma^2,
\label{eq:quad}
\eeq
where $m_\sigma$ denotes the mass of $\sigma$~\footnote{
As mentioned before, the effective mass during inflation may be
different from that in the low energy. Our arguments remains intact as
long as the effective mass is smaller than the Hubble parameter.
}.  The symmetry of
$U(\sigma)$ is the $Z_2$ symmetry in this case.  If $m_\sigma$ is
larger than $H_I$, the fluctuations of $\sigma$ get suppressed, while
$\sigma$ acquires fluctuations if $m_\sigma < H_I$.  In the
following we focus on the case of $m_\sigma < H_I$.

We first give a rough sketch of how non-Gaussianity is generated in
the primordial curvature fluctuations.  During inflation, the scalar
field $\sigma$ likely sits at the origin, $\sigma = 0$, based on the
symmetry argument. For the moment we assume this is the case, and we
will discuss this issue in detail in Sec.~\ref{sec:4}.  Since the mass
is lighter than the Hubble parameter, $\sigma$ acquires quantum
fluctuations, $\delta \sigma$, around the origin.  The magnitude of
$\delta \sigma$ is roughly equal to $H_I$.  After inflation, $\sigma$
starts to oscillate around the origin when the Hubble parameter
becomes comparable to $m_\sigma$, with an amplitude taking a different
value in each local patch of the universe.  The energy density of
$\sigma$ is proportional to the amplitude squared, $(\delta
\sigma)^2$~\footnote{
Precisely speaking, one has to take account of the spatially averaged
value of $\sigma$, which does not vanish in the observable finite
universe.  Then the dependence on $\delta \sigma$ becomes slightly
modified.  Nevertheless this intuitive explanation gives a
qualitatively valid picture.
}.  Taking $\delta \sigma$ as a Gaussian variable, the resultant
energy density exhibits strong deviation from the Gaussian
distribution~\cite{Linde:1996gt,Lyth:2006gd}, hence the name an
ungaussiton.  For simplicity, we assume that the ungaussiton $\sigma$
decays into radiation in the visible sector~\footnote{
The ungaussiton can generate sizable non-Gaussianity even if it does
not decay.  In this case, the non-Gaussianity arises from the
isocurvature fluctuations, which leave the characteristic signatures
in the CMB anisotropy~\cite{Kawasaki:2008sn}. Indeed, the $Z_2$
symmetry of the ungaussiton may naturally account for the stability to
become dark matter.
}.  Then, the fluctuations in the energy density of the ungaussiton
turn into small corrections to the adiabatic fluctuations arising from
the inflaton.  Although the corrections might be small in the
amplitude of the curvature perturbations, they can significantly
change higher order correlation funcitions such as the three-point
function which characterizes the strength of non-Gaussianty.

Let us now estimate non-Gaussianity produced by the ungaussiton.
First we define the fluctuations of $\sigma$, taking account of the
fact that our observable universe is finite. Assume that an ensemble
average of $\sigma$ vanishes. Since our observable universe is finite,
the spatial average of $\sigma$ over the comoving volume corresponding
to the present Hubble horizon does {\it not} vanish.  In order to
calculate the spatial average of $\sigma$, we need to define the time
slicing. For later convenience, we take a flat time slicing at $t=t_*$
when the smallest scale of interest leaves the horizon during
inflation.  Let us denote the spatial average of $\sigma$ on the flat
slicing by $\bar{\sigma}$ and define the fluctuation $\delta
\sigma_*({\vec x})$as $\delta \sigma_* ({\vec x}) \equiv \sigma
(t_*,{\vec x})-\bar{\sigma}$.  If the mass of $\sigma$ is much lighter
than $H_I$, those fluctuations defined on the slicing can be evaluated
when each scale leaves the horizon.  In the following we assume that
the inflation lasts long enough that $\sigma$ acquires an almost
scale-invariant Gaussian fluctuations which extend beyond the present
Hubble horizon. To put it more precisely, we define the fluctuations
in a box of a size $L$, which is (at least) several times larger than
the present Hubble horizon.  In fact, the fluctuations of the scales
beyond the current horizon size contribute to the non-vanishing
spatial average of $\sigma$.

According to the $\delta N$ formalism~\cite{Starobinsky:1986fxa,
  Sasaki:1995aw, Sasaki:1998ug, Lyth:2004gb}, on sufficiently large
scales, the curvature perturbation $\zeta$ on the uniform energy
density hypersurface at $t = t_f$ is equal to the perturbations in the
number of e-foldings between the uniform density slicing at $t = t_f$
and the initial flat slicing at $t=t_*$:
\beq
\zeta(t_f, {\vec x}) \;=\; N(t_f, t_*, {\vec x}) - \bar{N}(t_f, t_*)
\label{deltaN}
\eeq
with
\bea
N(t_f, t_*, {\vec x}) &=& \int_{t_*}^{t_f} H(t, {\vec x}) \,dt,\non\\
\bar{N}(t_f, t_*) &=& \int_{t_*}^{t_f} \bar{H}(t) \,dt,
\eea
where $N(t_f, t_*, {\vec x})$ is the number of e-foldings of expansion
between the two slicings, and $\bar{N}$ is that of the background
universe (or more precisely, the spatially averaged one).  Similarly,
$H(t,{\vec x})$ denotes the local
expansion rate, while $\bar{H}(t)$ denotes the background one.

We assume that the ungaussiton $\sigma$ decays after the
reheating~\footnote{ This is just for the sake of simplicity. If it is
  the inflaton that decays later, we should take $t_f$ after the
  reheating. However, the non-Gaussianity tends to be suppressed in
  this case, since the energy density of the ungaussiton becomes
  small.}, and set $t_f$ at the time well after the decay of $\sigma$.
As mentioned, we take $t_*$ at the time when the smallest scale of
interest crossed the Hubble horizon during inflation.  The curvature
perturbation $\zeta$ is conserved after $t=t_f$, that is, $\zeta$
becomes independent of $t_f$, since the adiabatic pressure condition
is satisfied.  Then we can regard the number of the $e$-foldings as
the function of $\phi_* \equiv \phi(t_*,{\vec x})$ and $\sigma_*
\equiv \sigma (t_*,{\vec x})$, and expand $\zeta$ in terms of the
fluctuations $\delta \phi_*$ and $\delta \sigma_*$, which are the
perturbations of $\phi_*$ and $\sigma_*$, as
\begin{eqnarray}
\zeta \;\approx\; N_\phi \delta \phi_*+N_\sigma \delta \sigma_*+
\frac{1}{2} N_{\sigma \sigma} \delta \sigma_*^2+\frac{1}{6} N_{\sigma
\sigma \sigma} \delta \sigma_*^3+\frac{1}{24}N_{\sigma \sigma \sigma \sigma} \delta \sigma_*^4
+\cdots,
\label{eq:zeta}
\end{eqnarray}
where the indices denote the partial derivative with respect to the
variables, i.e., $N_\phi=\partial N/\partial \phi,~N_\sigma= \partial
N/\partial \sigma$, etc..  Those coefficients depend on the background
evolution of the scalar fields $\phi$ and $\sigma$ from $t=t_*$ until
$t=t_f$. In particular, we use the spatially averaged value, $\sigma
=\bar{\sigma}$, as the initial condition of the background evolution.
Here and in what follows we omit the dependence on the comoving
coordinate ${\vec x}$ unless otherwise stated.  We assume that the
first term on the right hand side (RHS) of Eq.~(\ref{eq:zeta}), coming
from the inflaton fluctuations, dominates the two-point function of
$\zeta$.  As for the relevant higher order correlation functions, we
assume that the leading contributions come from the ungaussiton
fluctuations, and we have neglected second and higher order terms in
$\delta \phi_*$ in Eq.~(\ref{eq:zeta}).  It is known that the
non-Gaussianity of $\zeta$ coming from the intrinsic non-Gaussianity
of $\delta \phi_*$ and $\delta \sigma_*$ is far below the
observational sensitivity such as Planck~\cite{Maldacena:2002vr,
  Seery:2005gb, Seery:2006vu, Arroja:2008ga}.  Hence we can treat both
$\delta \phi_*$ and $ \delta \sigma_*$ as the Gaussian
variables~\footnote{Although we have defined $\delta \sigma$ as the
  deviation from $\bar{\sigma}$, one can still regard $\delta
  \sigma_*$ as the Gaussian fluctuations.  This is because the
  superhorizon fluctuations contribute to $\bar{\sigma}$, while
  $\delta \sigma$ represents fluctuations inside the horizon.}.

Before going further, it will be worth mentioning how the symmetry
comes into the game. The expression (\ref{eq:zeta}) itself is rather
generic, but, what is peculiar to the ungaussiton mechanism is the
relative size of the coefficients.  To see this, let us concentrate on
$N_\sigma$, which depends on the background evolution of $\sigma$. As
mentioned above, we take $\sigma = \bar{\sigma}$ as the initial
condition. Let us expand $N_\sigma$ around another initial condition,
$\sigma = 0$.
\beq \left.N_\sigma \right|_{\bar{\sigma}} \;=\; \left.N_\sigma
\right|_{0} + \left.N_{\sigma \sigma}\right|_{0} \bar{\sigma} +
\frac{1}{2}\left.N_{\sigma \sigma \sigma}\right|_{0} \bar{\sigma}^2+
\cdots, \eeq
where we have explicitly shown the initial values of $\sigma$.  What
the $Z_2$ symmetry tells us is that the potential, and therefore the
background evolution, is even around the origin.  That is to say, the
functions $N_{\sigma}$ and $N_{\sigma \sigma \sigma}$ should vanish
when $\sigma = 0$ is used as the initial condition.  That is,
\beq
\left.N_\sigma \right|_{0} \; = \; 0,~~~\left.N_{\sigma \sigma \sigma}\right|_{0} = 0.
\eeq
Therefore the expansion of $\left.N_\sigma \right|_{\bar{\sigma}}$
around $\sigma = 0$ actually starts with a term linear in
$\bar{\sigma}$.  Noting that $\bar{\sigma}$ is actually comparable to
the fluctuations $\delta \sigma_*$, one can see that the term linear
in $\delta \sigma_*$ in (\ref{eq:zeta}) is comparable to the term
quadratic in $\delta \sigma_*$. Intuitively speaking, since $\sigma$
stays at the origin based on the symmetry arguments, if one expands a
function of $\sigma$, $f(\sigma)$, around the origin, the leading term
should start with $\sigma^2$, as long as $f(\sigma)$ satisfies
$f(\sigma) = f(-\sigma)$ in the interested range of $\sigma$.  When
one takes account of the finiteness of the current horizon scale, this
feature appears in the suppression of
$\left.N_\sigma\right|_{\bar{\sigma}}$ as shown above. Let us
emphasize again that the suppression is due to the presence of the
symmetry, and that this makes the higher order term quadratic in
$\delta \sigma_*$ very important.  For the same reason, the cubic term
of $\delta \sigma_*$ is comparable to the quartic one, and therefore
we keep the quartic term of $\delta \sigma_*$ for the moment.

The power spectrum $P_\zeta$, bispectrum $B_\zeta$ and trispectrum
$T_\zeta$ of the curvature perturbations are defined as
\begin{eqnarray}
&&\langle \zeta_{\vec k_1} \zeta_{\vec k_2} \rangle_c = {( 2\pi )}^3
\delta ({\vec k_1}+{\vec k_2}) \,P_\zeta (k_1), \\ &&\langle
\zeta_{\vec k_1} \zeta_{\vec k_2} \zeta_{\vec k_3} \rangle_c = {( 2\pi
)}^3 \delta ({\vec k_1}+{\vec k_2}+{\vec k_3})\, B_\zeta (k_1, k_2,
k_3), \\ &&\langle \zeta_{\vec k_1} \zeta_{\vec k_2} \zeta_{\vec k_3}
\zeta_{\vec k_4} \rangle_c= {( 2\pi )}^3 \delta ({\vec k_1}+{\vec
k_2}+{\vec k_3}+{\vec k_4} ) \, T_\zeta ({\vec k_1}, {\vec k_2}, {\vec
k_3}, {\vec k_4} ),
\end{eqnarray}
where $\zeta_{\vec k}$ are Fourier components of $\zeta$, i.e.,
$\zeta_{\vec k} \equiv \int d^3 x e^{-i {\vec k} \cdot {\vec x}} \zeta
({\vec x})$, $k_i \equiv |{\vec k}_i|$ for $i = 1 \cdots 4$, and the
subscript $c$ means that we take the connected part of the
corresponding correlator. It is useful to define ${\cal P}_\zeta
\equiv k^3/(2\pi^2)\, P_\zeta$.  Let us also define the power spectra
of $\delta \phi$ and $\delta \sigma$,
\bea
\la \delta \phi_{\vec k_1} \delta \phi_{\vec k_2} \ra
&=& (2\pi)^3 \delta({\vec k_1}+{\vec k_2}) P_\phi(k_1),\\
\la \delta \sigma_{\vec k_1} \delta \sigma_{\vec k_2} \ra
&=& (2\pi)^3 \delta({\vec k_1}+{\vec k_2}) P_\sigma(k_1),\\
{\cal P}_{\phi(\sigma)}&\equiv& \frac{k^3}{2\pi^2} P_{\phi(\sigma)}.
\eea
Neglecting the tilt of the power spectra, we have
\beq
{\cal P}_\phi \simeq {\cal P}_\sigma \approx \lrfp{H_I}{2 \pi}{2}.
\eeq
As mentioned above, the inflaton dominates the two-point correlator of
$\zeta$, i.e.,
\beq
P_\zeta \;\simeq\; N_\phi^2 P_\phi. 
\eeq
As for the bispectrum, the dominant contributions come from the
ungaussiton, which, up to sixth order in $\delta \sigma_*$, gives
\begin{eqnarray}
B_\zeta (k_1, k_2, k_3) &\simeq& \frac{N_\sigma^2 N_{\sigma \sigma}}{N_\phi^4} 
\left( P_\zeta (k_1) P_\zeta(k_2)+2~{\rm perms.} \right) \nonumber \\
&&+\,\frac{N_{\sigma \sigma}^3}{ N_\phi^6 } \int \frac{d^3 {\vec q}}{(2\pi)^3 } \,
P_\zeta(q) P_\zeta(|{\vec k_1}+{\vec q}|) P_\zeta(|{\vec k_2}-{\vec q}|)
 \nonumber \\
&&+\,\frac{N_\sigma N_{\sigma \sigma} N_{\sigma \sigma \sigma}}{2N_\phi^6}
 \left(  \int \frac{d^3 {\vec q}}{(2 \pi)^3} \,P_\zeta(q) P_\zeta(|{\vec k_2}-{\vec q}|) 
 P_\zeta(k_1)   +5~{\rm perms.} \right),
 \label{bispectrum}
\end{eqnarray}
where the third term was not taken into account in
Refs.~\cite{Lyth:2006gd, Lyth:2005fi, Boubekeur:2005fj, Lyth:2007jh}.
The first term is quartic order in $\delta \sigma_*$, while the
remaining terms containing the momentum integral are sixth order in
$\delta \sigma_*$.

In a similar way, the trispectrum up to eighth order in $\delta
\sigma_*$ can be written as
\begin{eqnarray}
T_\zeta ({\vec k_1}, {\vec k_2}, {\vec k_3}, {\vec k_4} )&\simeq&
 \frac{N_\sigma^2 N_{\sigma \sigma}^2}{N_\phi^6} \left( P_\zeta (k_1) P_\zeta (k_2) P_\zeta (k_{13})+11~{\rm perms.} \right) \nonumber \\
&&+\, \frac{N_\sigma^3 N_{\sigma \sigma \sigma}}{N_\phi^6} \left( P_\zeta (k_1) P_\zeta (k_2) P_\zeta (k_3)+3~{\rm perms.} \right) \nonumber \\
&&+\,\frac{N_{\sigma \sigma}^4}{N_\phi^8} 
\left(\int \frac{d^3{\vec q}}{{(2\pi)}^3}\,
 P_\zeta(q) P_\zeta(|{\vec k_1}-{\vec q}|) P_\zeta(|{\vec k_2}+{\vec q}|)P_\zeta( |{\vec k_1}+{\vec k_3}-{\vec q}|)+2~{\rm perms.} \right) \nonumber \\
&&+\,\frac{N_\sigma N_{\sigma \sigma}^2 N_{\sigma \sigma \sigma}}{ N_\phi^8} \left( 
\int \frac{d^3{\vec q}}{{(2\pi)}^3}\,
 P_\zeta(q) P_\zeta(|{\vec k_2}+{\vec q}|) P_\zeta( |{\vec k_1}+{\vec k_4}-{\vec q}|) P_\zeta(k_1) 
 +11~{\rm perms.} \right) \nonumber \\
&&+\,\frac{N_\sigma N_{\sigma \sigma}^2 N_{\sigma \sigma \sigma}}{N_\phi^8} \left( 
\int \frac{d^3{\vec q}}{(2\pi)^3}\, P_\zeta(q) P_\zeta(|{\vec k_3}+{\vec q}|)
P_\zeta(k_1) P_\zeta(k_{12})
+11~{\rm perms.} \right) \nonumber \\
&&+\,\frac{N_{\sigma}^2 N_{\sigma \sigma \sigma}^2}{2 N_\phi^8}
 \left( \int \frac{d^3{\vec q}}{(2\pi)^3}\, 
 P_\zeta(q) P_\zeta(|{\vec k_1}+{\vec k_3}+{\vec q}|)P_\zeta(k_1) P_\zeta(k_{2})
 +11~{\rm perms.} \right) 
 \nonumber \\
&&+\,\frac{N_\sigma^2 N_{\sigma \sigma} N_{\sigma \sigma \sigma \sigma}}{2 N_\phi^8} 
\left(\int \frac{d^3{\vec q}}{(2\pi)^3}\, P_\zeta(q) P_\zeta(|{\vec k_3}-{\vec q}|) 
 P_\zeta(k_1) P_\zeta(k_2) +11~{\rm perms.} \right), \label{trispectrum}
\end{eqnarray}
where $k_{ij} \equiv |{\vec k_i}+{\vec k_j}|$. We have here dropped
small logarithmic corrections proportional to $\log(L/R)$, where $R$
denotes a smoothing scale.  The first three terms agree with those in
\cite{Lyth:2006gd, Boubekeur:2005fj, Lyth:2007jh}.

It is conventional to express $B_\zeta$ and $T_\zeta$ in terms of the
non-linearity parameters $f_{\rm NL},~\tau_{\rm NL}$ and $g_{\rm NL}$
defined by~\cite{Byrnes:2006vq}
\begin{eqnarray}
&&B_\zeta (k_1,k_2,k_3)=\frac{6}{5} f_{\rm NL} ( P_\zeta (k_1) P_\zeta
(k_2)+2~{\rm perms.} ), \label{bi1}\\ 
&&T_\zeta ({\vec k_1},{\vec k_2},{\vec
k_3},{\vec k_4})=\tau_{\rm NL} ( P_\zeta (k_{13})
P_\zeta (k_3) P_\zeta (k_4)+11~{\rm perms.}) \nonumber \\
&&\hspace{35mm}+\frac{54}{25} g_{\rm NL} ( P_\zeta (k_1) P_\zeta (k_2)
P_\zeta (k_3)+3~{\rm perms.}). \label{tri1}
\end{eqnarray}
If we include the terms containing the momentum integral in $B_\zeta$
and $T_\zeta$, the parametrization of Eqs.~(\ref{bi1}) and
(\ref{tri1}) does not correctly represent the bispectrum and the
trispectrum.  However, if we consider some limiting configurations of
the wavenumber vectors, so-called ``squeezed" configurations, we can
approximately express $B_\zeta$ and $T_\zeta$ in terms of $f_{\rm
  NL},~\tau_{\rm NL}$ and $g_{\rm NL}$.

Let us here make another approximation which drastically simplifies
our arguments. In the ungaussiton scenario, it turns out that those
terms containing $N_{\sigma \sigma \sigma}$ or $N_{\sigma \sigma
  \sigma \sigma}$ in $T_\zeta$ are suppressed compared to the terms
containing only $N_\sigma$ and $N_{\sigma \sigma}$, by the ratio of
the ungaussiton energy density to the total one at the time of its
decay.  The ratio must be smaller than $\sim 10^{-5}$ not to exceed
the primordial fluctuations produced from the inflaton.  Hence if some
quantity of interest starts from terms which contain only $N_\sigma$
and/or $N_{\sigma \sigma}$, we will drop higher order terms including
$N_{\sigma \sigma \sigma}$ or $N_{\sigma \sigma \sigma \sigma}$.

In the case of the bispectrum, the squeezed configuration is such that
one of $\{k_i\}$ with $i = 1,2,3$ is much smaller than the length of
the other two wavevectors; e.g., $k_1 \ll k_2, k_3$.  For such
configuration, $f_{\rm NL}$ can be written as~\cite{Lyth:2006gd,
Lyth:2005fi, Boubekeur:2005fj, Lyth:2007jh}
\begin{eqnarray}
\frac{6}{5}f_{\rm NL} \simeq \frac{1}{N_\phi^4} \left( N_\sigma^2 N_{\sigma \sigma}+N_{\sigma \sigma}^3 {\cal P}_\sigma \log (k_b L) \right), \label{eq:fnl}
\end{eqnarray}
where $k_b \equiv {\rm min}\{k_i\} \,(i=1,2,3)$ and $L$ is the size of
a box in which the perturbations are defined. 
Since we are interested in the perturbation in the observable
universe,
$L$ should be taken to be ${\cal O}(1/H_0)$,
where $H_0$ is the present value of the Hubble parameter.

As for the trispectrum, the situation becomes a littile bit more
complicated.  If the four wavenumber vectors are such that ${\rm
min}\{ k_i, |{\vec k_j}+{\vec k_\ell}| \}$ is much smaller than the
other elements of $\{ k_i,|{\vec k_j}+{\vec k_\ell}| \}$, where
$(i,j,\ell)=\{1,2,3,4\}$, $\tau_{\rm NL}$ can be expressed
by~\cite{Lyth:2006gd, Boubekeur:2005fj, Lyth:2007jh}
\begin{eqnarray}
\tau_{\rm NL} \simeq \frac{1}{N_\phi^6} \left( N_\sigma^2 N_{\sigma \sigma}^2+N_{\sigma \sigma}^4 {\cal P}_\sigma \log (k_t L) \right), \label{eq:tnl}
\end{eqnarray}
where $k_t \equiv {\rm min}\{ k_i, |{\vec k_j}+{\vec k_\ell}| \}$.
Meanwhile, if the configuration of the four wavenumber vectors is such
that two of them is much smaller than the other two, $g_{\rm NL}$ is
given by
\begin{eqnarray}
\frac{54}{25}g_{\rm NL}\simeq \frac{1}{N_\phi^6} \left( N_\sigma^3 N_{\sigma \sigma \sigma}+3N_\sigma^2 N_{\sigma \sigma} N_{\sigma \sigma \sigma \sigma} {\cal P}_\sigma \log (k_{g1} L) +3 N_\sigma N_{\sigma \sigma}^2 N_{\sigma \sigma \sigma} {\cal P}_\sigma \log (k_{g2} L) \right),
\label{eq:gnl}
\end{eqnarray}
where $k_{g1}^3 \equiv {\rm min}\{ {k_i k_j k_\ell} \}$ and $k_{g2}^3
\equiv {\rm min}\{ k_i^2 k_j \}$.  Note that $g_{\rm NL}$ starts from
the terms linear in $N_{\sigma \sigma \sigma}$ or $N_{\sigma \sigma
\sigma \sigma}$.  Hence the trispectrum is dominated by $\tau_{\rm
NL}$ terms if the leading non-Gaussianity is generated by the
ungaussiton.  Here and in what follows we neglect $g_{\rm NL}$ terms
and the tilt of the power spectra, for simplicity.

The non-linearity parameters depend on the thermal history of the
universe after inflation, since the coefficients such as $N_\sigma$
and $N_{\sigma\sigma}$ depend on the evolution of the inflaton and the
ungaussiton.  We consider a case that the ungaussiton decays after the
reheating, since we are interested in the case that the
non-Gaussianity becomes large. That is to say, we focus on the case of
\beq
\Gamma_\sigma \;<\; \Gamma_\phi,
\eeq
where $\Gamma_{\sigma(\phi)}$ is the decay rate of the ungaussiton
(inflaton).  There are still two cases to be considered, depending on
whether the ungaussiton starts oscillations before or after the
reheating.

If the ungaussiton starts its oscillations after the inflaton decay,
or equivalently, if $m_\sigma < \Gamma_\phi$, a relevant part of the
e-folding number $N$ that depends on $\sigma$ is given
by~\cite{Ichikawa:2008iq}
\begin{eqnarray}
N(t_f,t_*)\;\supset\;\frac{1}{24} \sqrt{\frac{\pi}{2}} \alpha^2 p+{\cal O}(p^2),
\label{nsigma}
\end{eqnarray}
where $\alpha =2\sqrt{2/\pi}\, \Gamma(5/4) \approx 1.45$ is a
numerical constant and $p \equiv ({\bar
\sigma}/M_P)^2\sqrt{m_\sigma/\Gamma_\sigma}$.  The parameter $p$
roughly represents the ratio of the ungaussiton energy density to the
total one at the time when the ungaussiton decays.  If the ungaussiton
dominates the universe before its decay, the resultant curvature
perturbations would become highly non-Gaussian, which is inconsistent
with the current observations.  Hence we require $p \ll 1$.  Note that
we have assumed that the curvature perturbation dominantly comes from
the inflaton, while the ungaussiton gives only small correction to
it. This actually gives a severer bound on $p$, i.e., $p \ll
O(10^{-5})$.  Substituting (\ref{nsigma}) into (\ref{eq:fnl}) and
(\ref{eq:tnl}), we obtain
\begin{eqnarray}
&&\frac{6}{5}f_{\rm NL}=\frac{\alpha^6}{216} {\left( \frac{\pi}{2} \right)}^{3/2} {\cal P}_\zeta \epsilon^3 {\left( \frac{\Gamma_\sigma}{m_\sigma} \right)}^{-3/2} \bigg \{ {\left( \frac{\bar \sigma}{H_I/2\pi} \right)}^2+\log (k_b L) \bigg \}, \label{fnl}\\ 
&&\tau_{\rm NL}= \frac{\pi^2 \alpha^8}{5184} {\cal P}_\zeta \epsilon^4 {\left( \frac{\Gamma_\sigma}{m_\sigma} \right)}^{-2} \bigg \{ {\left( \frac{\bar \sigma}{H_I/2\pi} \right)}^2+ \log (k_t L) \bigg \}, \label{taunl} 
\end{eqnarray}
where $\epsilon$ is one of the slow-roll parameters defined by
$\epsilon \equiv M_P^2/2\, |V_\phi/V|^2$ ($V$ is the scalar potential
of the inflaton), and it is related to $N_\phi$ and ${\cal P}_\zeta$
as
\beq
N_\phi\;=\;\frac{1}{\sqrt{2\epsilon} M_P},~~~~~{\cal P}_\zeta\;=\;\frac{H_I^2}{8 \pi^2 \epsilon M_P^2}.
\eeq
The WMAP normalization gives ${\cal P}_\zeta \approx 2\times 10^{-9}$~\cite{Komatsu:2008hk}.  

 Let us here discuss the magnitudes of the terms in the curly
  bracket of (\ref{fnl}) and (\ref{taunl}).  Since we are interested
  in a scalar fluctuating around the origin, the first term should be
  $O(1)$.  There are several ways to realize such situation as stated
  in Sec.~\ref{sec:1}.  In order to estimate the magnitude of the
  second term, we need to specify values of $k$ and $L$.  What values
  of $L$ should be taken depends on the spatial size of the
  observational data. For instance, suppose that we have observational
  data on a variable $X$ in a small patch of the sky.  Then we should
  take $L$ to be the size of the patch.  We can define a homogeneous
  part of $X$ by taking the spatial average over the region.  The
  fluctuations is obtained by subtracting
  the homogeneous mode from $X$. Those fluctuations with the
  wavelengths larger than $L$ cannot be ditinguished with the
  homogeneous part, and therefore they are regarded as the homogeneous
  mode.  In our case, since we are interested in the fluctuations of
  the scales between $k \sim H_0$ and $k \sim 10^{4}H_0$, we should
  take $L \sim {\cal O}(H_0^{-1})$, and the second term is $O(1)$ for
  interested ranges of the scales.  Therefore, both terms in the curly
  brackets are ${\cal O}(1)$.

The fact that both terms in the curly brackets are ${\cal O}(1)$ means
that if the ungaussiton is responsible for the non-Gaussianity of the curvature
perturbations, the non-linearity parameters $f_{\rm NL}$ and
$\tau_{\rm NL}$  have the logarithmic dependence on the spatial
scales~\footnote{ Note that we have neglected the scale dependence of
${\cal P}_\zeta$.  However, it can be determined by observations, and
so, in principle one can extract the scale dependence of the
non-linearity parameters.}.  In particular, they increase
logarithmically as one goes to small scales.

When does the non-Gaussianity become large?  The
condition, $f_{\rm NL}>f_{\rm NL}^{\rm min}$, is met if
\begin{eqnarray}
\epsilon^3 {\left( \frac{\Gamma_\sigma}{m_\sigma} \right)}^{-3/2} \gtrsim 7\times 10^{10}
\left( \frac{f_{\rm NL}^{\rm min}}{10} \right),
\label{large1}
\end{eqnarray}
where we have set the terms in the curly brackets on RHS of
(\ref{fnl}) and (\ref{taunl}) to be unity. Equivalenty, we can express
the same condition in terms of the Hubble parameter during inflation
instead of the slow-roll parameter $\epsilon$;
\begin{eqnarray}
H_I \;\gtrsim\; 6\times 10^{12} ~{\rm GeV} ~{\left( \frac{g_{*d}}{g_{*{\rm osc}}} \right)}^{1/8} {\left( \frac{T_d}{1~{\rm GeV}}\right)}^{1/2} {\left( \frac{T_{\rm osc}}{10^8~{\rm GeV}} \right)}^{-1/2} {\left( \frac{f_{\rm NL}^{\rm min}}{10} \right)}^{1/6}, \label{condition1}
\end{eqnarray}
where $T_d$ and $T_{\rm osc}$ are the temperatures of the radiation
when the ungaussiton decays and when the ungaussiton starts to
oscillate, respectively.  Here $g_{*d}$ and $g_{*{\rm osc}}$ are the
effective number of light degrees freedom at $T=T_d$ and $T_{\rm
osc}$, respectively.

Let us next consider the other case that the ungaussiton starts its
oscillations before the inflaton decay.  In this case, we have only to
replace $m_\sigma$ with $\Gamma_\phi$ in Eqs.~(\ref{large1}) and
(\ref{condition1}), employing the sudden decay approximation on the
inflaton decay.  Hence the condition, $f_{\rm NL} > f_{\rm NL}^{\rm
min}$, can be written as
\begin{eqnarray}
\epsilon^3 {\left( \frac{\Gamma_\sigma}{\Gamma_\phi} \right)}^{-3/2} \;\gtrsim\; 
7\times 10^{10}\left( \frac{f_{\rm NL}^{\rm min}}{10} \right),
\end{eqnarray}
or equivalently,
\begin{eqnarray}
H_I \;\gtrsim\; 6\times 10^{12} ~{\rm GeV} {\left( \frac{g_{*d}}{g_{*R}} \right)}^{1/8} {\left( \frac{T_d}{1~{\rm GeV}} \right)}^{1/2}{\left( \frac{T_R}{10^8~{\rm GeV}} \right)}^{-1/2}{\left( \frac{f_{\rm NL}^{\rm min}}{10} \right)}^{1/6}, \label{condition2}
\end{eqnarray}
where $T_R$ is the reheating temperature of the inflaton, and $g_{*R}$
the effective number of light degree of freedom at $T=T_R$.

There is the upper bound on the Hubble parameter during inflation as
$H_I \lesssim 10^{14} {\rm GeV}$ from the WMAP
result~\cite{Komatsu:2008hk}.  Meanwhile, both (\ref{condition1}) and
(\ref{condition2}) give the lower bounds on $H_I$.  Notice that both
the conditions (\ref{condition1}) and (\ref{condition2}) require the
decay temperature be much lower than the reheating temperature, i.e.,
the life time of the ungaussiton must be rather long.  The longevity
is necessary for the ungaussiton to give non-negligible contribution
to the energy density, and therefore to the non-Gaussianity.  It
suggests that the ungaussiton must couple only weakly to the visible
sector. Suppose that $\sigma$ couples to the visible sector strongly.
Then $\sigma$ can be immediately thermalized when the universe is
reheated by the inflaton decay, which would significantly suppress the
non-Gaussianity.  Thus the ungaussiton must be indeed weakly coupled
to the visible sector.

\begin{figure}[t]
\begin{center}
\includegraphics[width=110mm]{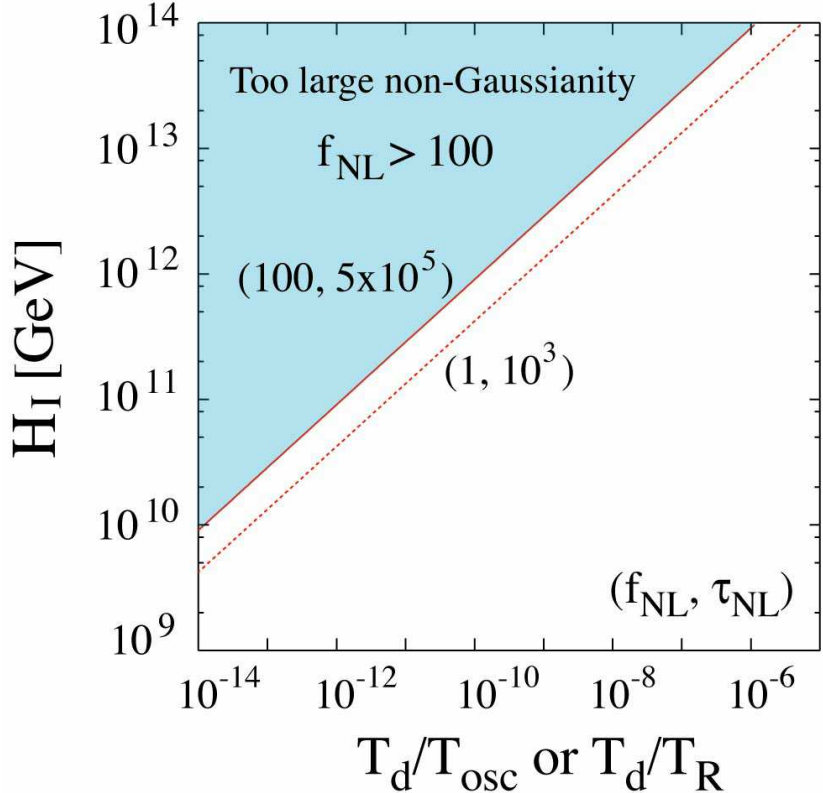}
\caption{Contours of the non-linearity parameters $f_{\rm NL}$ and $\tau_{\rm NL}$;
$(f_{\rm NL},\tau_{\rm NL}) = (100, 5\times 10^5)$ (solid line) and $(1,10^3)$ (dotted line).
Shaded region is already excluded by the WMAP 5yr data.
Here we set $g_{*{\rm osc}}=g_{*d}=g_{*R}=100$ and
${\bar \sigma}=H_I/2\pi$, and drop the logarithmic dependence.}
\label{fig:fnl}
\end{center}
\end{figure}

In Fig.~\ref{fig:fnl}, we show the contours of the non-linearity
parameters $f_{\rm NL}$ and $\tau_{\rm NL}$ as functions of
$T_d/T_{\rm osc}$ (or $T_d/T_R$) and $H_I$.  In the figure, we set
$g_{*{\rm osc}}=g_{*d}=g_{*R}=100$ and ${\bar \sigma}=H_I/2\pi$, and drop the
logarithmic dependence for simplicity.  We see that the region of
large $H_I$ and small $T_d/T_{\rm osc}$ (or $T_d/T_R$) is already
excluded by the observational constraint, $|f_{\rm NL}| \lesssim
100$~\cite{Komatsu:2008hk}.  We also find that the slopes of the
contours do not depend on the non-linearity parameters: $H_I \propto
\sqrt{T_d/T_{\rm osc}}$ or$\sqrt{T_d/T_R}$ for all the non-linearity
parameters.  This is because both $f_{\rm NL}$ and $\tau_{\rm NL}$
depend only on the combination $\epsilon
\sqrt{m_\sigma/\Gamma_\sigma}$ (or $\epsilon
\sqrt{\Gamma_\phi/\Gamma_\sigma}$), as can be seen from
Eqs.~(\ref{fnl}) and (\ref{taunl}).

The fact that all the non-linearity parameters depend only on one
combination of the model parameters means that we have an universal
relation between $f_{\rm NL}$ and $\tau_{\rm NL}$, which is independent
of the model parameters.  Such a consistency relation is given by
\begin{eqnarray}
\tau_{\rm NL}&=& C\, {\cal P}_\zeta^{-\frac{1}{3}} \left(\frac{6}{5} f_{\rm NL}\right)^{\frac{4}{3}}, \nonumber \\
&\approx& 1\times 10^3 \,C\, f_{\rm NL}^{4/3}, \label{consis1}
\end{eqnarray}
where we have used ${\cal P}_\zeta \simeq 2 \times
10^{-9}$~\cite{Komatsu:2008hk}.  Here $C$ is a numerical coefficient
of order unity defined by
\begin{eqnarray}
C &\equiv&
\ds{ \frac{ \frac{{\bar \sigma}^2}{{(H_I/2\pi)}^2}+ \log k_t L }{ {\left( \frac{{\bar \sigma}^2}{{(H_I/2\pi)}^2}+ \log k_b L \right)}^{4/3}}}.
\end{eqnarray}

As a remarkable fact, when $f_{\rm NL} > 1$, the trispectrum of the
curvature perturbations becomes inevitably very large $\tau_{\rm NL}
\gg 1$.  Although the current bound on $\tau_{\rm NL}$ is very weak,
we will have much stronger bound in the near future: e.g., $|\tau_{\rm
NL}| \lesssim 560$ for Planck~\cite{Kogo:2006kh}.  Hence if the
non-Gaussianity $f_{\rm NL} > 1$ is generated by the ungaussiton
mechanism, we should also see the strong non-Gaussianity through the
trispectrum, which will be quite useful to distinguish the ungaussiton
scenario from the others that predict different consistency relations
between the bispectrum and the trispectrum~\cite{Suyama:2007bg,
Buchbinder:2007at, Ichikawa:2008iq}.

Let us here show that the consistency relation (\ref {consis1}) can be
understood in an intuitive way.  The bispecturm is roughly given by
$\la \zeta^3 \ra \sim N_{\sigma \sigma}^3 \la (\delta \sigma) ^6\ra$,
while the trispectrum is $\la \zeta^4 \ra \sim N_{\sigma \sigma}^4 \la
(\delta \sigma) ^8\ra$.  Therefore, as long as the quadratic term $
(\delta \sigma)^2$ is the dominant source of the non-Gaussianity, we
naively expect $\la \zeta^4 \ra \sim (\la \zeta^3 \ra)^{4/3}$, which
correctly reproduces $\tau_{\rm NL} \sim {\cal P_\zeta}^{-1/3}f_{\rm
NL}^{4/3} \sim 10^3 f_{\rm NL}^{4/3}$.  We can easily extend this
argument to the $n$-th correlator. The leading $n$-th non-linearity
parameter defined in a similar fashion, $f_ {\rm NL}^{(n)}$, is
roughly equal to ${\cal P_\zeta}^{(3-n)/3} f_{\rm NL}^{n/3} \sim
10^{3(n-3)} f_{\rm NL}^{n/3}$.  One can see that this simple
argument breaks down if $N_ \sigma$ is not suppressed as in the
curvaton scenario, since the bispectrum would be dominated by $\la
\zeta^3 \ra \sim N_\sigma^2 N_ {\sigma \sigma} \la (\delta \sigma) ^4
\ra$.  In that case there are different consistency relations~\cite
{Ichikawa:2008iq}.  Thus it is very important to test the consistency
relations by observations, in order to distinguish different scenarios
to produce non-Gaussianity.

As mentioned before, there must be hierarchy between $T_R$ (or $T_{\rm
  osc}$) and $T_d$. Since the ungaussiton must decay before the big
bang nucleosynthesis (BBN), $T_d$ must be higher than
$10\,$MeV~\cite{Kawasaki:1999na}.  Then, the reheating temperature is
bounded below: e.g., $T_R \gtrsim 10^2\,$GeV for $H_I = 10^{14}$\,GeV,
and $T_R \gtrsim 10^{9}$GeV for $H_I = 10^{11}$\,GeV, in order to have
$f_{\rm NL} \gtrsim 1$.  Such high reheating temperature as $T_R
\gtrsim 10^{9}$GeV may lead to the overproduction of the
gravitinos~\cite{Weinberg:zq,Krauss:1983ik}, when the local SUSY is
assumed.  One way to ameliorate the tension is to consider the heavy
or ultralight gravitino mass: $m_{3/2} \gtrsim O(10)$\,TeV or $m_{3/2}
\lesssim O(10)$\,eV.  Another is to introduce a late-time entropy
production after the ungaussiton
decays~\cite{Lyth:1995ka,Kawasaki:2004rx}.  Note that the density
fluctuations after the ungaussiton decays are adiabatic, although they
contain sizable non-Gaussianity.  Once it becomes adiabatic, the
non-Gaussianity will be inherent in the subsequent evolution of the
universe, whatever it is like. In particular, the non-Gaussianity does
not change even if there is late-time entropy production after the
ungaussiton decays, unless isocurvature perturbations exist in the
fields that induce entropy production.

Let us here make a comment on the isocurvature perturbations.  If the
baryon number or the dark matter abundance is fixed before the
undaussiton decays, the non-Gaussian isocurvature perturbations are
generated. The amplitude of the isocurvature perturbations are smaller
than the current bounds, for the non-Gaussianity with $f_{\rm NL} <
100$. Although we do not consider its effects throughout this paper,
the isocurvature perturbations, if observed, may provide another clue
on the ungaussiton mechanism.

Before closing this section, we note that the symmetries does not have
to be exact to realize the ungaussiton scenario.  The explicit
breaking of the symmetries will generically makes the potential
minimum deviate from the origin, and $\bar{\sigma}$ may have a large
value.  However, if the deviation is the order of $H_I$, this is
indistinguishable from the non-vanishing $\bar{\sigma}$ due to the
fluctuations on large scales beyond the size of the observable
universe.  Also, as in the case of an axion field with a shift
symmetry: $a \rightarrow a+\alpha$, the enhanced-symmetry point may
not exist. Even in this case, if the initial position of the scalar
happens to be near the minimum, it can be an ungaussiton if it is
light during inflation.  Therefore our discussions above remain intact
in these cases.

%%%%%%%%%%%%%%%%%%%%%%%%%%%%%%%
\section{Models} 
\label{sec:3}
%%%%%%%%%%%%%%%%%%%%%%%%%%%%%%%

Let us here give some models to realize our idea. We will give
detailed analysis on each model in the coming paper~\cite{st}.  One
example is the modulus field, which appears in the
supergravity/superstring theory and it has a very flat potential. If
the position of the modulus field during inflation is deviated from
the minimum in the low energy, it leads to the cosmological moduli
problem, since the large energy density stored in the modulus field
can easily modify and spoil the standard evolution of the universe.
One of the solutions to the moduli problem is to presume that the
moduli fields having symmetry-enhanced points in their field space
stay at the points during and after inflation~\cite{Dine:1998qr,
Kofman:2004yc,Watson:2004aq,Cremonini:2006sx,
Greene:2007sa},
which would greatly suppress the cosmological abundance of the
moduli. Such moduli fields become ungaussitons, if they are light
during inflation~\footnote{ The abundance of the such modulus field
was investigated in Ref.~\cite{Felder:1999wt} to study how the moduli
problem is relaxed by the mechanism.}. Since the moduli must decay
before BBN starts, the modulus mass must be larger than
$O(10^2)$\,TeV. Assuming that the modulus has couplings with the
standard-model particles suppressed by the Planck scale, the decay
temperature is given by around $O(100)\,$MeV for the modulus mass of
$10^3\,$TeV.  From Fig.~\ref{fig:fnl}, one can see that large
non-Gaussianity is generated for e.g. $H_I = 10^{11}\,$GeV and $T_R =
10^{10}\,$GeV.  In the case of the moduli, the needed hierarchy
between $T_R$ and $T_d$ is realized due to the weakness of the modulus
couplings.

Another is a right-handed sneutrino, $\tilde{N}$, a superpartner of
the right-handed neutrino, $N$. The right-handed sneutrino is odd
under the $R$-parity, which determines the origin of $\tilde{N}$.  If
the mass of $\tilde{N}$ is light during inflation, it can be a
ungaussiton. If the neutrino Yukawa coupling $y$ is small enough and
if the mass of $\tilde{N}$ is relatively light, the decay of
$\tilde{N}$ can be delayed.  Also one can neglect the thermal
production of $\tilde {N}$ for sufficiently small Yukawa couplings.
Therefore, it is relatively easy to satisfy the relation
(\ref{condition2}) if the model is close to the case of the Dirac
neutrino.  For instance, the decay temperature is roughly estimated by
\beq 
T_d\;\sim\;0.1\,y\,\sqrt{m_{\tilde N} M_P}, 
\eeq 
where $m_{\tilde N}$ is the mass of ${\tilde N}$.  Therefore, large
non-Gaussianity is generated for e.g. $y=10^{-6}$, $m_{\tilde N} =
1$\,TeV, $H_I = 10^{14}$\,GeV and $T_R = 10^{7}$\,GeV.  Note that the
gravitationally produced right-handed sneutrino cannot explain the
baryon asymmetry through leptogenesis~\cite{Fukugita:1986hr}, since the baryon asymmetry
would have too large isocurvature perturbations.

The last example is a flat direction of the supersymmetric standard
model (SSM).  There are many flat directions in SSM, and they are
parametrized by a gauge-invariant monomial such as $udd$ and
$eLL$. The flat directions are composed of squarks, sleptons and
Higgs, and therefore, the origins of the flat directions are
determined by the SM gauge symmetry.  In addition, we expect that some
of the flat directions may be light during inflation, since there are
many. However, since they interact with the standard-model particle
through the SM gauge interactions, they might be easily thermalized
after the reheating. Interestingly, there is one possibility:
$Q$-balls~\cite{Coleman:1985ki}.  After the flat direction starts
oscillating, it generically experiences spatial instabilities and
deforms into non-topological solitons called $Q$-balls~\footnote{Note
that the spatial scale of the $Q$-balls are much smaller than the
cosmological scales of our interest.}.  Once the $Q$-balls are formed,
their life time is typically very long since the decay and evaporation
processes occur only around the surfaces of the
$Q$-balls~\cite{Qdecay}.  The relation (\ref{condition2}) can be
satisfied for certain values of the inflation scale and the reheating
temperature.  For instance, we take $H_I = 10^{14}$\,GeV.  Assuming
the gravity mediation or the anomaly-mediation~\cite{AMSB}, the
typical charge of the $Q$-ball (of the gravity-mediaiton type) is $Q
\sim 10^{19}$~\cite{KK2}.  Such $Q$-balls decay or evaporate depending
on the details of the flat directions, but the typical decay or
evaporation temperature is between ${\cal O}(1)$\,GeV to ${\cal
O}(100)$\,GeV.  Then the relation (\ref{condition2}) is met for $T_R
\simeq 10^7 \sim 10^9$\,GeV.  What makes this scenario slightly
complicated is that the decay temperature is related to the charge of
the $Q$-balls, which in turn depends on the initial
amplitude. Therefore, in this model using the $Q$-balls, we also have
to take account of the non-Gaussianity produced by the so-called
modulated reheating scenario. Although we expect that the results
derived in the previous section can be qualitatively valid even in
this case, there must be quantitative difference. The detailed
analysis of this scenario is beyond the scope of this paper, and we
leave it for the future work.  Nevertheless we would like to emphasize
that the flat directions in SSM can be ungaussitons, which makes us to
believe that the ungaussiton mechanism is indeed feasible and that one
can find further explicit models.

Each scenario described above has its own implications on the
inflationary scale, the mass scale and the lifetime of the
ungaussiton. We will discuss this issue in a separate paper.

%%%%%%%%%%%%%%%%%%%%%%%%%%%%%%%
\section{Discussion and Conclusions} 
\label{sec:4}
%%%%%%%%%%%%%%%%%%%%%%%%%%%%%%%

We would like to emphasize here that the ungaussiton is different from
a curvaton~\cite{curvaton}, which has been extensively studied as a
possible source for the non-Gaussianity~\cite{Lyth:2002my,
Bartolo:2003jx, Bartolo:2004if, Enqvist:2005pg, Lyth:2006gd,
Malik:2006pm, Sasaki:2006kq, Huang:2008ze}.  The original motivation
of the curvaton scenario was to explain the adiabatic density
fluctuations by the fluctuations in the curvaton field.  Therefore it
is crucial for the curvaton to develop a large non-zero expectation
value during inflation. The large expectation value helps the curvaton
to dominate the energy density of the universe, or at least it makes
the energy density of the curvaton sizable if it does not dominate.
At the same time, the fluctuations in the energy density of the
curvaton field becomes almost Gaussian with relatively small
non-Gaussianity. On the other hand, the ungaussiton has a classically
vanishing field value during inflation.  Therefore its energy density
exhibits strong non-Gaussianity, and it is not responsible for the
almost Gaussian curvature perturbations observed in the CMB. Note that
this difference results in the smaller number of the parameters in the
ungaussiton scenario relative to the curvaton scenario, which makes
the former more predictive. This is because the amplitude of the
ungaussiton is determined solely by the Hubble parameter during
inflation, while the amplitude of the curvaton is a free parameter in
the curvaton scenario~\footnote{In the ungaussiton scenario, there is
numerical uncertainty of order unity in the effective amplitude of
$\sigma$, since it arises from the purely quantum fluctuations. Still,
we have less number of the dimensionful parameters.  }. Thanks to this
feature, we have obtained the consistency relation between the
bispectrum and the trispectrum (\ref{consis1}).  Although the
techniques to calculate the non-Gaussianity are common to the case of
the curvaton, the background philosophy is different: the presence of
the symmetry-enhanced points in the field space is the key idea for
our arguments.

 Lastly, let us discuss how the ungaussiton can remain around the
  minimum of the potential during inflation. Since the mass is assumed
  to be lighter than the Hubble parameter, one may wonder why the
  ungaussiton can find its origin and sit there during and after
  inflation. Our answer is as follows.

There are several ways to infer how the initial position is
determined, although we cannot decisively predict it unless the
dynamics before (or during) inflation is unravelled.  If the position
is determined in a probabilistic way, the probability distribution
must have its extremum at the origin, if the theory respects the
symmetry around the origin. (This probability distribution has nothing
to do with the quantum fluctuations.  The initial condition we are
talking here is about the position before the ungaussiton starts to
fluctuate.)  We do not know if the extremum is either (local) minimum
or maximum. If the probability becomes maximum at the origin, the
scalar field likely sits at the origin at the beginning of the
universe.  Such scalar can be an ungaussiton.  On the other hand, the
symmetry would be spontaneously broken if the probability is minimum
at the origin, which is more suitable for the curvaton scenario.

One can also imagine that the initial position is determined by some
dynamics.  For instance, if there is another inflationary phase before
the primordial inflation, the scalar field may acquire a positive mass
of the order of the Hubble parameter around the origin. Then the
scalar will settle down at the origin during the pre-inflation.  Or,
there might be a radiation-dominated phase after the pre-inflation and
before the last inflation starts. If the ungaussiton is thermalized
during this phase, it will stay near the origin when the last
inflation starts.

If the initial position is somehow set around the origin as described
above, the ungaussiton will start to fluctuate about the origin during
the last inflation.  If it lasted for $50 - 60$ e-foldings, it still
remains around the origin.  However, if the last inflation lasted for
much longer period, the situation changes.

As the inflation lasts longer, the fluctuations will be accumulated,
and the initial condition set before the inflation becomes less
important, and finally, it will be forgotten in the end.  The
asymptotic distribution is known as the BD distribution.  If it is
reached, the ungaussiton takes a value of ${\cal O} (H_I^2/m)$ in most
regions of the whole universe. Our observable universe may happen to
be in a (small) region where the ungaussiton takes a value of
$O(H_I)$.  This is a fine-tuning if $m \ll H_I$, since such regions
are quite rare~\cite{Linde:1996gt}.  However, this is not the case if
the effective mass $m$ is not much smaller than $H_I$.  If we take
$m=0.1 H_I$, for instance, the RMS of the field values is ${\cal O}(10
H_I)$.  Therefore, we can find a (sub)universe where the spatial
average of the ungaussiton field is $O(H_I)$ without severe
fine-tunings.  Note that the fluctuations of the ungaussiton in the
observable universe are not damped significantly, for $m \lesssim
0.1H_I$, because the suppression factor is given 
by $\sim \exp({-\frac{m^2}{H_I^2} N_*})$
where $N_*$ is the e-folding number from the time when the mode crossed 
the horizon to the time of the inflation end.
Such possibility can be naturally realized in the context of
supergravity.  In supergravity, a scalar field generically acquires a
mass of $O(H_I)$ during inflation.  Depending on a precise form of the
Kahler potential, the effective mass can be lighter than $H_I$.  (One
can even make the correction vanish if the Kahler potential is finely
tuned.)  Therefore, without severe fine-tunings, it is possible that
the effective mass is only slightly smaller than $H_I$.

After all, we cannot tell the initial position of a scalar field due
to our ignorance regarding the scalar dynamics in the inflationary
phase.  However, without any theoretical prejudice, we can still
expect that a scalar field possibly sits at the origin, simply because
the origin is so special in its field space.  

If there are the symmetry-enhanced points, we believe it plausible
that the scalar fields stay at the points during and after inflation.
If some of them are light during inflation, they acquire quantum
fluctuations. Such light scalars are ungaussitons: their energy
densities are necessarily highly non-Gaussian!  Importantly, such
non-Gaussianity must have been imprinted in the CMB spectrum. For the
effects to be large enough to be actually measured by the ongoing and
planned observations, the relations Eqs.~(\ref{condition1}) or
(\ref{condition2}) must be satisfied.  Note that the criterion for a
scalar field to be an ungaussiton is not strict at all. Any scalar
field charged under some symmetries can become an ungaussiton, if it
is light during inflation.  If such scalars are mediocre in nature, we
may well expect that there are indeed such ungaussitons satisfying the
relations Eq.~(\ref{condition1}) or (\ref{condition2}).

We do not claim that the ungaussiton mechanism is superior to the
curvaton scenario. These two are independent and interesting
possibilities to generate sizable non-Gaussianity in the CMB spectrum,
and as we have mentioned before, the ungaussiton scenario predicts
distinctive bispectrum and trispectrum: the enhancement at the small
scales and the consistency relation (\ref{consis1}).  
Therefore, the future observations may be able to
tell one from the other.

\bigskip
\bigskip

\noindent {\bf Acknowledgments:} F.T. would like to thank Matt Buckley
and Sourav Mandal for suggesting the name of the particle.
We would also like to thank M. Kawasaki for comments.  
This work was supported by World Premier International Research 
Center Initiative (WPI Program), MEXT, Japan.


\begin{thebibliography}{100}

\bibitem{string}
See for example, J.~Polchinski, {\em String Theory},
\newblock Cambridge Monographs on Mathematical Physics, 1998.
 
\bibitem{ModuliProblem}
%
%\cite{Coughlan:1983ci}
%\bibitem{Coughlan:1983ci}
  G.~D.~Coughlan et al
  %, W.~Fischler, E.~W.~Kolb, S.~Raby and G.~G.~Ross,
  %``Cosmological Problems For The Polonyi Potential,''
  Phys.\ Lett.\ B {\bf 131} (1983) 59;
  %%CITATION = PHLTA,B131,59;%%
%\cite{Banks:1993en}
%\bibitem{Banks:1993en}
  T.~Banks, D.~B.~Kaplan and A.~E.~Nelson,
  %``Cosmological implications of dynamical supersymmetry breaking,''
  Phys.\ Rev.\ D {\bf 49} (1994) 779;
%  [arXiv:hep-ph/9308292];\\
  %%CITATION = HEP-PH 9308292;%%
  %%Cited 53 times in SPIRES-HEP
%\cite{deCarlos:1993jw}
%\bibitem{deCarlos:1993jw}
  B.~de Carlos et al
  %, J.~A.~Casas, F.~Quevedo and E.~Roulet,
  %``Model independent properties and cosmological implications of the dilaton
  %and moduli sectors of 4-d strings,''
  Phys.\ Lett.\ B {\bf 318} (1993) 447.
%  [arXiv:hep-ph/9308325].
  %%CITATION = HEP-PH 9308325;%%
  %%Cited 44 times in SPIRES-HEP 
 
\bibitem{MGP}
 %\cite{Endo:2006zj}
%\bibitem{Endo:2006zj}
  M.~Endo, K.~Hamaguchi and F.~Takahashi,
  %``Moduli-induced gravitino problem,''
  Phys.\ Rev.\ Lett.\  {\bf 96}, 211301 (2006);\\
%  [arXiv:hep-ph/0602061].
  %%CITATION = PRLTA,96,211301;%%
  %\cite{Nakamura:2006uc}
%\bibitem{Nakamura:2006uc}
  S.~Nakamura and M.~Yamaguchi,
  %``Gravitino production from heavy moduli decay and cosmological moduli
  %problem revived,''
  Phys.\ Lett.\  B {\bf 638}, 389 (2006).
 % [arXiv:hep-ph/0602081].
  %%CITATION = PHLTA,B638,389;%%
 
\bibitem{MGP2}
%\cite{Dine:2006ii}
%\bibitem{Dine:2006ii}
  M.~Dine, R.~Kitano, A.~Morisse and Y.~Shirman,
  %``Moduli decays and gravitinos,''
  Phys.\ Rev.\  D {\bf 73}, 123518 (2006).
%  [arXiv:hep-ph/0604140].
  %%CITATION = PHRVA,D73,123518;%%
 
%\cite{Endo:2006tf}
\bibitem{Endo:2006tf}
  M.~Endo, K.~Hamaguchi and F.~Takahashi,
  %``Moduli / inflaton mixing with supersymmetry breaking field,''
  Phys.\ Rev.\  D {\bf 74}, 023531 (2006).
%  [arXiv:hep-ph/0605091].
  %%CITATION = PHRVA,D74,023531;%% 
  
%\cite{Dine:1998qr}
\bibitem{Dine:1998qr}
  M.~Dine, Y.~Nir and Y.~Shadmi,
  %``Enhanced symmetries and the ground state of string theory,''
  Phys.\ Lett.\  B {\bf 438}, 61 (1998).
%  [arXiv:hep-th/9806124].
  %%CITATION = PHLTA,B438,61;%%  
  
 
%\cite{Kofman:2004yc}
\bibitem{Kofman:2004yc}
  L.~Kofman, A.~Linde, X.~Liu, A.~Maloney, L.~McAllister and E.~Silverstein,
  %``Beauty is attractive: Moduli trapping at enhanced symmetry points,''
  JHEP {\bf 0405}, 030 (2004)
  [arXiv:hep-th/0403001].
  %%CITATION = JHEPA,0405,030;%%

%\cite{Watson:2004aq}
\bibitem{Watson:2004aq}
  S.~Watson,
  %``Moduli stabilization with the string Higgs effect,''
  Phys.\ Rev.\  D {\bf 70}, 066005 (2004)
  [arXiv:hep-th/0404177].
  %%CITATION = PHRVA,D70,066005;%%  

%\cite{Cremonini:2006sx}
\bibitem{Cremonini:2006sx}
  S.~Cremonini and S.~Watson,
  %``Dilaton dynamics from production of tensionless membranes,''
  Phys.\ Rev.\  D {\bf 73}, 086007 (2006)
  [arXiv:hep-th/0601082].
  %%CITATION = PHRVA,D73,086007;%%    
  
 %\cite{Greene:2007sa}
\bibitem{Greene:2007sa}
  B.~Greene, S.~Judes, J.~Levin, S.~Watson and A.~Weltman,
  %``Cosmological Moduli Dynamics,''
  JHEP {\bf 0707}, 060 (2007)
  [arXiv:hep-th/0702220].
  %%CITATION = JHEPA,0707,060;%% 
 
  
\bibitem{GPP} 
%\cite{Zeldovich:1971mw}
%\bibitem{Zeldovich:1971mw}
  Y.~B.~Zeldovich and A.~A.~Starobinsky,
  %``Particle production and vacuum polarization in an anisotropic gravitational
  %field,''
  Sov.\ Phys.\ JETP {\bf 34}, 1159 (1972)
  [Zh.\ Eksp.\ Teor.\ Fiz.\  {\bf 61}, 2161 (1971)];
  %%CITATION = ZETFA,61,2161;%%
%\cite{Grishchuk:1975uf}
%\bibitem{Grishchuk:1975uf}
  L.~P.~Grishchuk,
  %``The Amplification Of Gravitational Waves And Creation Of Gravitons In The
  %Isotropic Universe,''
  Lett.\ Nuovo Cim.\  {\bf 12}, 60 (1975)
  [Erratum-ibid.\  {\bf 12}, 432 (1975)];
  %%CITATION = NCLTA,12,60;%%
%\cite{Mamaev:1976zb}
%\bibitem{Mamaev:1976zb}
  S.~G.~Mamaev, V.~M.~Mostepanenko and A.~A.~Starobinsky,
  %``Particle Creation From Vacuum Near An Homogeneous Isotropic Singularity,''
  Zh.\ Eksp.\ Teor.\ Fiz.\  {\bf 70} (1976) 1577;
  %%CITATION = ZETFA,70,1577;%%
%\cite{Grib:1976pw}
%\bibitem{Grib:1976pw}
  A.~A.~Grib, S.~G.~Mamaev and V.~M.~Mostepanenko,
  %``Particle Creation From Vacuum In Homogeneous Isotropic Models Of The
  %Universe,''
  Gen.\ Rel.\ Grav.\  {\bf 7} (1976) 535.
  %%CITATION = GRGVA,7,535;%%

%\cite{Linde:1996gt}
\bibitem{Linde:1996gt}
  A.~D.~Linde and V.~F.~Mukhanov,
  %``Nongaussian isocurvature perturbations from inflation,''
  Phys.\ Rev.\  D {\bf 56}, 535 (1997)
  [arXiv:astro-ph/9610219];
  %%CITATION = PHRVA,D56,535;%%
%\cite{Linde:2005yw}
%\bibitem{Linde:2005yw}
  A.~Linde and V.~Mukhanov,
  %``The curvaton web,''
  JCAP {\bf 0604}, 009 (2006)
  [arXiv:astro-ph/0511736].
  %%CITATION = JCAPA,0604,009;%%

%\cite{Durrer:1998sv}
\bibitem{Durrer:1998sv}
  R.~Durrer, M.~Gasperini, M.~Sakellariadou and G.~Veneziano,
  %``Seeds of large-scale anisotropy in string cosmology,''
  Phys.\ Rev.\  D {\bf 59}, 043511 (1999)
  [arXiv:gr-qc/9804076];
  %%CITATION = PHRVA,D59,043511;%%
%\cite{Durrer:1998zj}
%\bibitem{Durrer:1998zj}
%  R.~Durrer, M.~Gasperini, M.~Sakellariadou and G.~Veneziano,
  %``Massless (pseudo-)scalar seeds of CMB anisotropy,''
  Phys.\ Lett.\  B {\bf 436}, 66 (1998)
  [arXiv:astro-ph/9806015];
  %%CITATION = PHLTA,B436,66;%%
%\cite{Gasperini:1998bm}
%\bibitem{Gasperini:1998bm}
  M.~Gasperini and G.~Veneziano,
  %``Constraints on pre-big bang models for seeding large-scale anisotropy  by
  %massive Kalb-Ramond axions,''
  Phys.\ Rev.\  D {\bf 59}, 043503 (1999)
  [arXiv:hep-ph/9806327].
  %%CITATION = PHRVA,D59,043503;%%  

%\cite{Bunch:1978yq}
\bibitem{Bunch:1978yq}
  T.~S.~Bunch and P.~C.~W.~Davies,
  %``Quantum Field Theory In De Sitter Space: Renormalization By Point
  %Splitting,''
  Proc.\ Roy.\ Soc.\ Lond.\  A {\bf 360}, 117 (1978).
  %%CITATION = PRSLA,A360,117;%%


  %\cite{Yadav:2007yy}
\bibitem{Yadav:2007yy}
  A.~P.~S.~Yadav and B.~D.~Wandelt,
  %``Detection of primordial non-Gaussianity (fNL) in the WMAP 3-year data at
  %above 99.5% confidence,''
  arXiv:0712.1148 [astro-ph].
  %%CITATION = ARXIV:0712.1148;%%
 
%\cite{Komatsu:2008hk}
\bibitem{Komatsu:2008hk}
  E.~Komatsu {\it et al.}  [WMAP Collaboration],
  %``Five-Year Wilkinson Microwave Anisotropy Probe (WMAP\altaffilmark 1 )
  %Observations:Cosmological Interpretation,''
  arXiv:0803.0547 [astro-ph].
  %%CITATION = ARXIV:0803.0547;%% 

%\cite{Lyth:2006gd}
\bibitem{Lyth:2006gd}
  D.~H.~Lyth,
  %``Non-gaussianity and cosmic uncertainty in curvaton-type models,''
  JCAP {\bf 0606}, 015 (2006)
  [arXiv:astro-ph/0602285].
  %%CITATION = JCAPA,0606,015;%%

%\cite{Kawasaki:2008sn}
\bibitem{Kawasaki:2008sn}
  M.~Kawasaki, K.~Nakayama, T.~Sekiguchi, T.~Suyama and F.~Takahashi,
  %``Non-Gaussianity from isocurvature perturbations,''
  arXiv:0808.0009 [astro-ph].
  %%CITATION = ARXIV:0808.0009;%%

%\cite{Starobinsky:1986fxa}
\bibitem{Starobinsky:1986fxa}
  A.~A.~Starobinsky,
  %``Multicomponent de Sitter (Inflationary) Stages and the Generation of
  %Perturbations,''
  JETP Lett.\  {\bf 42} (1985) 152
  [Pisma Zh.\ Eksp.\ Teor.\ Fiz.\  {\bf 42} (1985) 124].
  %%CITATION = ZFPRA,42,124;%%

%\cite{Sasaki:1995aw}
\bibitem{Sasaki:1995aw}
  M.~Sasaki and E.~D.~Stewart,
  %``A General Analytic Formula For The Spectral Index Of The Density
  %Perturbations Produced During Inflation,''
  Prog.\ Theor.\ Phys.\  {\bf 95}, 71 (1996).
%  [arXiv:astro-ph/9507001].
  %%CITATION = PTPKA,95,71;%%

%\cite{Sasaki:1998ug}
\bibitem{Sasaki:1998ug}
  M.~Sasaki and T.~Tanaka,
  %``Super-horizon scale dynamics of multi-scalar inflation,''
  Prog.\ Theor.\ Phys.\  {\bf 99}, 763 (1998)
  [arXiv:gr-qc/9801017].
  %%CITATION = GR-QC 9801017;%%  

%\cite{Lyth:2004gb}
\bibitem{Lyth:2004gb}
  D.~H.~Lyth, K.~A.~Malik and M.~Sasaki,
  %``A general proof of the conservation of the curvature perturbation,''
  JCAP {\bf 0505}, 004 (2005)
  [arXiv:astro-ph/0411220].
  %%CITATION = JCAPA,0505,004;%%

%\cite{Maldacena:2002vr}
\bibitem{Maldacena:2002vr}
  J.~M.~Maldacena,
  %``Non-Gaussian features of primordial fluctuations in single field
  %inflationary models,''
  JHEP {\bf 0305}, 013 (2003)
  [arXiv:astro-ph/0210603].
  %%CITATION = JHEPA,0305,013;%%

%\cite{Seery:2005gb}
\bibitem{Seery:2005gb}
  D.~Seery and J.~E.~Lidsey,
  %``Primordial non-gaussianities from multiple-field inflation,''
  JCAP {\bf 0509}, 011 (2005)
  [arXiv:astro-ph/0506056].
  %%CITATION = JCAPA,0509,011;%%

%\cite{Seery:2006vu}
\bibitem{Seery:2006vu}
  D.~Seery, J.~E.~Lidsey and M.~S.~Sloth,
  %``The inflationary trispectrum,''
  JCAP {\bf 0701}, 027 (2007)
  [arXiv:astro-ph/0610210].
  %%CITATION = JCAPA,0701,027;%%  

%\cite{Arroja:2008ga}
\bibitem{Arroja:2008ga}
  F.~Arroja and K.~Koyama,
  %``Non-gaussianity from the trispectrum in general single field inflation,''
  arXiv:0802.1167 [hep-th].
  %%CITATION = ARXIV:0802.1167;%%  

%\cite{Byrnes:2006vq}
\bibitem{Byrnes:2006vq}
  C.~T.~Byrnes, M.~Sasaki and D.~Wands,
  %``The primordial trispectrum from inflation,''
  Phys.\ Rev.\  D {\bf 74}, 123519 (2006)
  [arXiv:astro-ph/0611075].
  %%CITATION = PHRVA,D74,123519;%%

%\cite{Lyth:2005fi}
\bibitem{Lyth:2005fi}
  D.~H.~Lyth and Y.~Rodriguez,
  %``The inflationary prediction for primordial non-gaussianity,''
  Phys.\ Rev.\ Lett.\  {\bf 95}, 121302 (2005)
  [arXiv:astro-ph/0504045].
  %%CITATION = PRLTA,95,121302;%%

%\cite{Boubekeur:2005fj}
\bibitem{Boubekeur:2005fj}
  L.~Boubekeur and D.~H.~Lyth,
  %``Detecting a small perturbation through its non-Gaussianity,''
  Phys.\ Rev.\  D {\bf 73}, 021301 (2006)
  [arXiv:astro-ph/0504046].
  %%CITATION = PHRVA,D73,021301;%%

%\cite{Lyth:2007jh}
\bibitem{Lyth:2007jh}
  D.~H.~Lyth,
  %``The curvature perturbation in a box,''
  JCAP {\bf 0712}, 016 (2007)
  [arXiv:0707.0361 [astro-ph]].
  %%CITATION = JCAPA,0712,016;%%

%\cite{Ichikawa:2008iq}
\bibitem{Ichikawa:2008iq}
  K.~Ichikawa, T.~Suyama, T.~Takahashi and M.~Yamaguchi,
  %``Non-Gaussianity, Spectral Index and Tensor Modes in Mixed Inflaton and
  %Curvaton Models,''
  arXiv:0802.4138 [astro-ph].
  %%CITATION = ARXIV:0802.4138;%%

%\cite{Kogo:2006kh}
\bibitem{Kogo:2006kh}
  N.~Kogo and E.~Komatsu,
  %``Angular Trispectrum of CMB Temperature Anisotropy from Primordial
  %Non-Gaussianity with the Full Radiation Transfer Function,''
  Phys.\ Rev.\  D {\bf 73}, 083007 (2006)
  [arXiv:astro-ph/0602099].
  %%CITATION = PHRVA,D73,083007;%%

%\cite{Suyama:2007bg}
\bibitem{Suyama:2007bg}
  T.~Suyama and M.~Yamaguchi,
  %``Non-Gaussianity in the modulated reheating scenario,''
  Phys.\ Rev.\  D {\bf 77}, 023505 (2008)
  [arXiv:0709.2545 [astro-ph]].
  %%CITATION = PHRVA,D77,023505;%%

%\cite{Buchbinder:2007at}
\bibitem{Buchbinder:2007at}
  E.~I.~Buchbinder, J.~Khoury and B.~A.~Ovrut,
  %``Non-Gaussianities in New Ekpyrotic Cosmology,''
  arXiv:0710.5172 [hep-th].
  %%CITATION = ARXIV:0710.5172;%%  
 
  \bibitem{Kawasaki:1999na}
  M.~Kawasaki, K.~Kohri and N.~Sugiyama,
  %``Cosmological Constraints on Late-time Entropy Production,''
  Phys.\ Rev.\ Lett.\  {\bf 82}, 4168 (1999)
  [arXiv:astro-ph/9811437];
  %%CITATION = ASTRO-PH 9811437;%%
  %\cite{Kawasaki:2000en}
%\bibitem{Kawasaki:2000en}
  %M.~Kawasaki, K.~Kohri and N.~Sugiyama,
  %``MeV-scale reheating temperature and thermalization of neutrino
  %background,''
  Phys.\ Rev.\ D {\bf 62}, 023506 (2000)
  [arXiv:astro-ph/0002127];
  %%CITATION = ASTRO-PH 0002127;%%
  S.~Hannestad,
%``What is the lowest possible reheating temperature?,''
Phys.\ Rev.\ D {\bf 70}, 043506 (2004)
[arXiv:astro-ph/0403291];
%%CITATION = ASTRO-PH 0403291;%%
  %\cite{Ichikawa:2005vw}
%\bibitem{Ichikawa:2005vw}
  K.~Ichikawa, M.~Kawasaki and F.~Takahashi,
  %``The oscillation effects on thermalization of the neutrinos in the  universe
  %with low reheating temperature,''
  Phys.\ Rev.\ D {\bf 72}, 043522 (2005)
  [arXiv:astro-ph/0505395].
  %%CITATION = ASTRO-PH 0505395;%% 
 
  \bibitem{Weinberg:zq}
    S.~Weinberg,
    Phys.\ Rev.\ Lett.\  {\bf 48}, 1303 (1982).
    %%CITATION = PRLTA,48,1303;%
    
\bibitem{Krauss:1983ik}
    L.~M.~Krauss,
    Nucl.\ Phys.\ B {\bf 227}, 556 (1983).
    %%CITATION = NUPHA,B227,556;%%  

%\cite{Lyth:1995ka}
\bibitem{Lyth:1995ka}
  D.~H.~Lyth and E.~D.~Stewart,
  %``Thermal inflation and the moduli problem,''
  Phys.\ Rev.\ D {\bf 53}, 1784 (1996).
  [arXiv:hep-ph/9510204].
 %%CITATION = HEP-PH 9510204;%%
%\cite{Kawasaki:2004rx}
\bibitem{Kawasaki:2004rx}
  M.~Kawasaki and F.~Takahashi,
  %``Late-time entropy production due to the decay of domain walls,''
  Phys.\ Lett.\ B {\bf 618}, 1 (2005)
  [arXiv:hep-ph/0410158].
  %%CITATION = HEP-PH 0410158;%%

\bibitem{st}
T.~Suyama and F.~Takahashi,
in preparation.
 
%\cite{Felder:1999wt}
\bibitem{Felder:1999wt}
  G.~N.~Felder, L.~Kofman and A.~D.~Linde,
  %``Gravitational particle production and the moduli problem,''
  JHEP {\bf 0002}, 027 (2000).
%  [arXiv:hep-ph/9909508].
  %%CITATION = JHEPA,0002,027;%% 
  
 %\cite{Fukugita:1986hr}
\bibitem{Fukugita:1986hr}
  M.~Fukugita and T.~Yanagida,
  %``Baryogenesis Without Grand Unification,''
  Phys.\ Lett.\  B {\bf 174}, 45 (1986).
  %%CITATION = PHLTA,B174,45;%% 
  
\bibitem{Coleman:1985ki}
  S.~R.~Coleman,
  %``Q Balls,''
  Nucl.\ Phys.\ B {\bf 262}, 263 (1985)
  [Erratum-ibid.\ B {\bf 269}, 744 (1986)].
  %%CITATION = NUPHA,B262,263;%%  

\bibitem{Qdecay}
A.~G.~Cohen, S.~R.~Coleman, H.~Georgi and A.~Manohar,
%``The Evaporation Of Q Balls,''
Nucl.\ Phys.\ B {\bf 272}, 301 (1986).
%%CITATION = NUPHA,B272,301;%%

   \bibitem{AMSB} 
 %\cite{Randall:1998uk}
%\bibitem{Randall:1998uk}
  L.~Randall and R.~Sundrum,
  %``Out of this world supersymmetry breaking,''
  Nucl.\ Phys.\ B {\bf 557}, 79 (1999);\\
%  [arXiv:hep-th/9810155].
  %%CITATION = HEP-TH 9810155;%%
 %\cite{Giudice:1998xp}
%\bibitem{Giudice:1998xp}
  G.~F.~Giudice, M.~A.~Luty, H.~Murayama and R.~Rattazzi,
  %``Gaugino mass without singlets,''
  JHEP {\bf 9812}, 027 (1998);\\
%  [arXiv:hep-ph/9810442].
  %%CITATION = HEP-PH 9810442;%% 
  %\cite{Bagger:1999rd}
%\bibitem{Bagger:1999rd}
  J.~A.~Bagger, T.~Moroi and E.~Poppitz,
  %``Anomaly mediation in supergravity theories,''
  JHEP {\bf 0004}, 009 (2000).
%  [arXiv:hep-th/9911029].
  %%CITATION = HEP-TH 9911029;%% 
 
\bibitem{KK2}
S.~Kasuya and M.~Kawasaki,
%``Q-ball formation in the gravity-mediated SUSY breaking scenario,''
Phys.\ Rev.\ D {\bf 62}, 023512 (2000).
%%CITATION = PHRVA,D62,023512;%%  
 
\bibitem{curvaton}
%\bibitem{Enqvist:2001zp}
K.~Enqvist and M.~S.~Sloth,
%``Adiabatic CMB perturbations in pre big bang string cosmology,''
Nucl.\ Phys.\ B {\bf 626}, 395 (2002);\\
%[arXiv:hep-ph/0109214]
%\bibitem{Lyth:2001nq}
D.~H.~Lyth and D.~Wands,
%``Generating the curvature perturbation without an inflaton,''
Phys.\ Lett.\ B {\bf 524}, 5 (2002);\\
%[arXiv:hep-ph/0110002];
%\bibitem{Moroi:2001ct}
T.~Moroi and T.~Takahashi,
%``Effects of cosmological moduli fields on cosmic microwave background,''
Phys.\ Lett.\ B {\bf 522}, 215 (2001)
[Erratum-ibid.\ B {\bf 539}, 303 (2002)].
%[arXiv:hep-ph/0110096].
%%%
 
%\cite{Lyth:2002my}
\bibitem{Lyth:2002my}
  D.~H.~Lyth, C.~Ungarelli and D.~Wands,
  %``The primordial density perturbation in the curvaton scenario,''
  Phys.\ Rev.\  D {\bf 67}, 023503 (2003)
  [arXiv:astro-ph/0208055].
  %%CITATION = PHRVA,D67,023503;%%

%\cite{Bartolo:2003jx}
\bibitem{Bartolo:2003jx}
  N.~Bartolo, S.~Matarrese and A.~Riotto,
  %``On non-Gaussianity in the curvaton scenario,''
  Phys.\ Rev.\  D {\bf 69}, 043503 (2004)
  [arXiv:hep-ph/0309033].
  %%CITATION = PHRVA,D69,043503;%%

%\cite{Bartolo:2004if}
\bibitem{Bartolo:2004if}
  N.~Bartolo, E.~Komatsu, S.~Matarrese and A.~Riotto,
  %``Non-Gaussianity from inflation: Theory and observations,''
  Phys.\ Rept.\  {\bf 402}, 103 (2004)
  [arXiv:astro-ph/0406398].
  %%CITATION = PRPLC,402,103;%%

%\cite{Enqvist:2005pg}
\bibitem{Enqvist:2005pg}
  K.~Enqvist and S.~Nurmi,
  %``Non-gaussianity in curvaton models with nearly quadratic potential,''
  JCAP {\bf 0510}, 013 (2005)
  [arXiv:astro-ph/0508573].
  %%CITATION = JCAPA,0510,013;%%

%\cite{Malik:2006pm}
\bibitem{Malik:2006pm}
  K.~A.~Malik and D.~H.~Lyth,
  %``A numerical study of non-gaussianity in the curvaton scenario,''
  JCAP {\bf 0609}, 008 (2006)
  [arXiv:astro-ph/0604387].
  %%CITATION = JCAPA,0609,008;%%

%\cite{Sasaki:2006kq}
\bibitem{Sasaki:2006kq}
  M.~Sasaki, J.~Valiviita and D.~Wands,
  %``Non-gaussianity of the primordial perturbation in the curvaton model,''
  Phys.\ Rev.\  D {\bf 74}, 103003 (2006)
  [arXiv:astro-ph/0607627].
  %%CITATION = PHRVA,D74,103003;%%

 %\cite{Huang:2008ze}
\bibitem{Huang:2008ze}
  Q.~G.~Huang,
  %``Large Non-Gaussianity Implication for Curvaton Scenario,''
  arXiv:0801.0467 [hep-th].
  %%CITATION = ARXIV:0801.0467;%% 
 
\end{thebibliography}
\end{document}